\documentclass[epj]{webofc}
\usepackage[varg]{txfonts}   
\usepackage{amsmath}
\usepackage{dsfont}
\usepackage{capt-of}
\usepackage[section]{placeins}
\usepackage[compact]{titlesec}
\wocname{EPJ Web of Conferences}
\woctitle{CONF12}
\renewcommand{\O}{\mathcal{O}}
\renewcommand{\l}{\ell}
\renewcommand{\d}{\mathrm{d}}

\newcommand{\mr}{m_{\mathrm{r}}}
\newcommand{\e}[1]{\mathrm{e}^{#1}}
\renewcommand{\ln}[1]{\mathrm{ln}\!\left({#1}\right)}
\newcommand{\lnSquare}[1]{\mathrm{ln}^2\!\left({#1}\right)}
\newcommand{\1}{\mathds{1}}
\renewcommand{\r}[2]{\rho_{#1,#2}}

\begin{document}
\selectlanguage{english}
\title{pNRQCD determination of E1 radiative transitions}

\author{Sebastian Steinbei{\ss}er\inst{1}\fnsep\thanks{\email{
sebastian.steinbeisser@tum.de}} \and Jorge Segovia\inst{1}}

\institute{Physik-Department, Technische Universit\"at M\"unchen,
James-Franck-Str.\,1, D-85748 Garching, Germany}

\abstract{This contribution contains the first numerical computation of the 
complete set of relativistic corrections of relative order $v^{2}$ for electric 
dipole (E1) transitions in heavy quarkonium; in particular, for the processes 
$\chi_{bJ}(1P) \to \Upsilon(1S) + \gamma$ with $J=0,\,1,\,2$. We assume 
that the momentum transfer of the heavy mesons involved in the reactions lies in 
the weak-coupling regime of the low-energy effective field theory potential 
non-relativistic QCD (pNRQCD) and thus a full perturbative calculation can be 
performed.}

\maketitle


\section{Introduction}
\label{sec:Introduction}

Electromagnetic transitions are often significant decay modes for bottomonium 
states below $B\bar{B}$ threshold ($10.56\,{\rm GeV}$), making them a suitable 
experimental tool to access the lowest spectra of bottomonia. For instance, 
the first $b\bar{b}$ states not directly produced in $e^{+}e^{-}$ 
collisions were the six triplet-$P$ states, $\chi_{b}(2P_{J})$ and 
$\chi_{b}(1P_{J})$ with $J=0,\,1,\,2$, discovered in radiative decays of the 
$\Upsilon(3S)$ and $\Upsilon(2S)$ in $1982$~\cite{Han:1982zk, Eigen:1982zm} and
$1983$~\cite{Klopfenstein:1983nx, Pauss1983439}, respectively.

One important feature of electromagnetic transitions is that they can be 
classified in a series of electric and magnetic multipoles. The most important 
ones are the E1 (electric dipole) and the M1 (magnetic dipole) transitions; 
higher order multipole modes E2, M2, E3, etc. appear in the spectrum, but 
since they are further suppressed one usually does not consider them. 
Processes involving electric dipole (E1) transitions happen more frequently 
than the ones induced by a magnetic dipole (M1). The branching fraction for E1 
transitions can indeed be significant for some lowest bottomonium states like 
the ones we shall study herein~\cite{Agashe:2014kda}: ${\cal B}(\chi_{b0}(1P)\to 
\Upsilon(1S)\gamma) = (1.76\pm0.35)\,\%$ (note that it is the largest exclusive 
branching fraction reported by the Particle Data Group 
(PDG)~\cite{Agashe:2014kda}), ${\cal B}(\chi_{b1}(1P)\to \Upsilon(1S)\gamma) = 
(33.9\pm2.2)\,\%$ and ${\cal B}(\chi_{b2}(1P)\to \Upsilon(1S)\gamma) = 
(19.1\pm1.2)\,\%$. 

Electric dipole (E1) transitions are defined through the property that they 
change the orbital angular momentum of the state by one unit, but not the spin. 
Therefore, the final state has different parity and C-parity than the initial 
one. Typical E1 quarkonium decays are the ones mentioned above: $2^{3}P_{J}\to 
1^{3}S_{1}+\gamma$. Here and in the following we denote the states as 
$n\,^{2s+1}\!\ell_{J}$, where $n=n_{r}+\ell+1$ corresponds to the principal 
quantum number with $n_{r}=0,\,1,\,\ldots$ the radial quantum number and $\ell$ 
the orbital angular momentum. The spin is denoted by $s$ and $J$ is the total 
angular momentum. 

The E1 (and M1) electromagnetic transitions have been treated for a long time 
by means of potential models that use non-relativistic reductions of QCD-based 
quark-antiquark interactions (see, {\it e.g.}, Ref.~\cite{Segovia:2016xqb} for a 
recent application to the bottomonium system). However, the progress made in 
effective field theories (EFTs) for studying heavy 
quarkonia~\cite{Brambilla:2004jw} and the new large set of accurate experimental 
data taken in the heavy quark sector by B-factories (BaBar, Belle and CLEO), 
$\tau$-charm facilities (CLEO-c, BESIII) and even proton-proton colliders (CDF, 
D0, LHCb, ATLAS, CMS) ask for a systematic and model-independent analysis (see, 
{\it e.g.}, Refs.~\cite{Brambilla:2010cs, Brambilla:2014jmp} for reviews).

Formulae and numerical treatment of M1 transitions within the effective field 
theory potential NRQCD (pNRQCD) can be found in Refs.~\cite{Brambilla:2005zw, 
Pineda:2013lta}. Therein, the relativistic corrections to the leading order (LO) 
expression (which counts as $k_{\gamma}^3/m^2$ where $k_{\gamma}$ is the photon 
energy) were computed in two different expansion schemes: (i) strict weak-coupling 
regime and (ii) including exactly the static potential in the LO Hamiltonian. 
Within the same theoretical framework, the corresponding formulae for E1 
transitions have been presented in Ref.~\cite{Brambilla:2012be}. In this case, the 
relativistic corrections to the LO decay width (that counts as 
$k_{\gamma}^{3}/(mv)^{2}$) are much more involved covering not only higher order 
terms in the E1 transition operator but also corrections to the initial and 
final wave function due to higher order potentials and higher order Fock states. 
These facts have hindered numerical computations of the E1 radiative decays 
within pNRQCD (for partial calculations see \cite{Pietrulewicz:2013ct}). This 
contribution aims to close this gap and calculate the decay rate of the reaction 
$2^3P_J \to 1^3S_1 + \gamma$ with $J=0,\,1,\,2$. As a first step, we shall assume 
that the soft scale lies in the strict weak-coupling regime of pNRQCD and thus a 
full perturbative calculation can be performed. These proceedings are based on the 
forthcoming publication \cite{steinbeisser}.


\section{Theoretical set-up}
\label{sec:TheoreticalFramework}

\subsection{Potential non-relativistic QCD (pNRQCD)}
\label{subsec:pNRQCD}

Heavy quarkonium systems are characterized by their non-relativistic nature, 
{\it i.e.}, the heavy quark bound-state velocity, $v$, satisfies $v\ll1$. This 
is reasonably fulfilled in bottomonium ($v^{2} \sim 0.1$) and to a certain 
extent in charmonium ($v^{2}\sim 0.3$). Moreover, at least, three widely 
separated scales appear: the heavy quark mass $m$ (hard scale), the relative 
momentum of the bound state $p\sim mv$ (soft scale) and the binding energy 
$E\sim mv^{2}$ (ultrasoft scale). With $v\ll1$, the following hierarchy of 
scales
\begin{equation}
\label{eq:hierarchy}
m \gg p \sim 1/r \sim mv  \gg E \sim mv^2\,
\end{equation}
is satisfied and this allows for a description in terms of EFTs for physical 
processes taking place at one of the lower scales. The integration out of modes 
associated with high-energy scales is performed as part of a matching procedure 
that enforces the equivalence between QCD and the EFT at a given order of the 
expansion in $v$. The final result is a factorization at the Lagrangian level 
between the high-energy modes, which are encoded in the matching coefficients, 
and the low-energy contributions carried by the dynamical degrees of freedom.

The suitable EFT to describe processes that take place at the scale $mv$ such 
as the E1 radiative transitions between the lowest heavy quarkonium states is 
potential NRQCD (pNRQCD)~\cite{Pineda:1997bj, Brambilla:1999xf}. It follows by 
integrating out the modes of order $p\sim 1/r \sim mv$ from 
NRQCD~\cite{Caswell:1985ui, Bodwin:1994jh} which in turn comes from QCD by 
integrating out the high energy modes of order $m$. Therefore, pNRQCD takes 
full advantage of the hierarchy of scales that appear in 
Eq.~(\ref{eq:hierarchy}), and makes a systematic and natural connection between 
quantum field theory and the Schr\"odinger equation. Schematically, the pNRQCD 
equation of motion takes the form
\begin{eqnarray*}
\,\left.
\begin{array}{ll}
&
\displaystyle{
\left[ i\partial_0-{\vec{p\,}^2 \over m} - V_s^{(0)}(r)\right] 
\phi(\vec{r},t,\vec{R}\,)=0} \\[0.4ex]
&
\displaystyle{\ + \ \mbox{corrections to the potential}} \\[0.4ex]
&
\displaystyle{\ +\ \mbox{interactions with other low-energy degrees of freedom}}
\end{array} \right\}
{\rm pNRQCD}
\end{eqnarray*}
where $V_s^{(0)}(r)$ is the static potential and $\phi(\vec{r}\,)$ is the 
$Q\bar{Q}$ field. Note here that the interactions with other low-energy degrees of 
freedom produce, among others, non-potential terms that account for singlet to 
octet transitions via ultrasoft gluons and provide loop corrections to the leading 
potential picture. Being induced by low-energy degrees of freedom they encode also 
non-perturbative effects.

The matching of pNRQCD depends on the relative size between the soft and the 
$\Lambda_{\rm QCD}$ scale. Two main situations can be distinguished, namely, the 
weak-coupling~\cite{Pineda:1997bj,Brambilla:1999xf} ($mv \gg \Lambda_{\rm QCD}$) 
and the strong-coupling~\cite{Brambilla:2000gk} ($mv \sim \Lambda_{\rm QCD}$) 
versions of pNRQCD. One major difference between them is that in the former the 
potential can be computed in perturbation theory unlike in the latter.

It is obvious that the weak-coupling version of pNRQCD is amenable for a 
theoretically much cleaner analysis. The observables can be computed as an 
expansion in $\alpha_s$ with increasing accuracy. Non-perturbative effects are 
suppressed by powers of $\Lambda_{\rm QCD}/(mv)$. Therefore, observables that 
could be computed with the weak-coupling version of pNRQCD are of the greatest 
interest.


\begin{figure}[!t]
\begin{center}
\sidecaption
\includegraphics[clip,width=0.50\textwidth,height=0.25\textheight]
{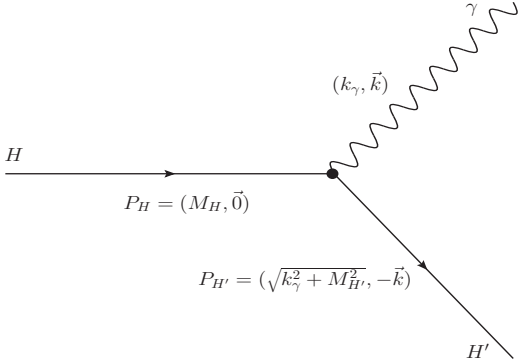} 
\caption{\label{fig:kinematics} Kinematics of the radiative transition $H\to 
H'\gamma$ in the rest frame of the initial-state quarkonium $H$, taken from \cite{Brambilla:2005zw}.}
\end{center}
\end{figure}

\subsection{Decay width of the $\mathbf{n^{3}P_{J} \to {n'}^3S_{1}\gamma}$ 
reaction}
\label{subsec:DecayWidth}

The complete decay rate $n^{3}P_{J}\to {n'}^3S_{1}\gamma$ reads up to order 
$k_{\gamma}^{3}/m^{2}$~\cite{Brambilla:2012be}
\begin{equation}
\begin{split}
\label{eq:FullDecayWidth}
&
\Gamma_{n^3P_J \to n'^3S_1\gamma} = \Gamma_{E1}^{(0)}\, \Bigg\{ 1 + R^{S=1}(J) 
- \frac{k_\gamma}{6m} - \frac{k_\gamma^2}{60} \frac{I_5^{(0)}(n1 \to 
n'0)}{I_3^{(0)}(n1 \to n'0)} \\
& 
+ \left[ \frac{J(J+1)}{2} - 2 \right] \Bigg[ -\left(1+\kappa_Q^{em}\right) 
\frac{k_\gamma}{2m} + \frac{1}{m^2} (1 + 2\kappa_Q^{em}) \frac{I_2^{(1)}(n1 \to 
n'0) + 2I_1^{(0)}(n1 \to n'0)}{I_3^{(0)}(n1 \to n'0)} \Bigg] \Bigg\} \,,
\end{split}
\end{equation}
where $R^{S=1}(J)$ includes the initial and final state corrections due to 
higher order potentials and higher order Fock states (see below). The remaining 
corrections within the brackets are the result of taking into account 
additional electromagnetic interaction terms in the Lagrangian suppressed by 
${\cal O}(v^2)$~\cite{Brambilla:2012be}. We have displayed terms proportional to 
the anomalous magnetic moment, $\kappa_{Q}^{\rm em}$, however these terms are at 
least suppressed by $\alpha_{s}(m)v^2$ and thus go beyond our accuracy and are therefore 
not considered in the numerical analysis. The LO decay width ($\sim 
k_{\gamma}^{3}/(mv)^2$) is
\begin{equation}
\label{eq:Gamma0}
\Gamma_{E1}^{(0)} = \frac{4}{9}\, \alpha_{em}\, e_Q^2\, k_\gamma^3 
\left[I_3^{(0)}(n1 \to n'0) \right]^{2} \,,
\end{equation}
with $\alpha_{em}$ the electromagnetic fine structure constant, $e_Q$ the charge 
of the heavy quarks in units of the electron charge, and $k_\gamma$ the photon 
energy determined by the kinematics shown in Fig.~\ref{fig:kinematics}:
\begin{equation}
k_{\gamma} = |\vec{k}| = \frac{M_{H}^{2}-M_{H'}^{2}}{2M_{H}} = (M_{H}-M_{H}') + 
{\cal O}\left(\frac{k_{\gamma}^{2}}{M_{H}}\right) \,.
\end{equation}
The function
\begin{equation}
I_N^{(k)}(n\l \to n'\l') = \int\limits_0^\infty \d r \, r^2 r^{N-2} 
R_{n'\l'}^{\ast}(r) \left[ \frac{\d^k}{\d r^k} R_{n\l}(r) \right]
\end{equation}
is a matrix element that involves the radial wave functions of the initial and 
final states. We shall assume that these states are solutions of the 
Schr\"odinger equation
\begin{equation}
\label{eq:SchEqu} 
H^{(0)} \psi_{n\l m}^{(0)}(\vec{r}\,) = E_n^{(0)} \psi_{n\l m}^{(0)}(\vec{r}\,) 
\,,
\end{equation}
with the leading order Hamiltonian in weakly-coupled pNRQCD given by
\begin{equation}
H^{(0)} = -\frac{\nabla^2}{2\mr} + V_s^{(0)}(r) = -\frac{\nabla^2}{2\mr} - C_F 
\frac{\alpha_s}{r} \,,
\end{equation}
where $C_F = 4/3$. Therefore, $\psi_{n\ell m}^{(0)}(\vec{r}\,)$ and $E_{n}^{(0)}$ 
can be written in the hydrogen-like form
\begin{align}
\psi_{n\l m}^{(0)}(\vec{r}\,) &= R_{nl}(r) Y_{\l m}(\Omega_r) = N_{n\l} \, 
\e{-\frac{\rho_n}{2}} \rho_n^\l \, L_{n-\l-1}^{2\l+1}(\rho_n) Y_{\l 
m}(\Omega_r) \,, \label{eq:eigenstate} \\
E_n^{(0)} &= -\frac{\mr C_F^2 \alpha_s^2}{2n^2} \,, \label{eq:eigenenergy}
\end{align}
where $\mr = m/2$ is the reduced mass of the $Q\bar{Q}$ system, $\rho_n = 2 r/ 
n a$ is a dimensionless variable with $a = 1/\mr C_F \alpha_s$ the 
Bohr radius. Finally, the normalization reads
\begin{equation}
N_{n\l} = \sqrt{\left( \frac{2}{n a} \right)^3 \frac{(n-\l-1)!}{2n[(n+\l)!]}} 
\,.
\end{equation}


\subsection{Relativistic wave-function corrections}
\label{subsec:RelativisticCorrections}

Due to higher order potentials and transitions between singlets and octets, the 
state in Eq.~(\ref{eq:eigenstate}) is not an eigenstate of the complete 
Hamiltonian. Therefore, one has to consider corrections to the wave function, 
which can contribute to the decay rate at the required order of precision 
($\sim\!\! k_{\gamma}^{3}/m^{2}$). To compute these corrections one applies the 
standard formalism of perturbation theory, either in the language of quantum 
mechanics or via Feynman diagrams.

\subsubsection{Corrections due to higher order potentials}

In order to account for corrections to the decay width of relative order 
$v^{2}$, we need to consider the complete Hamiltonian
\begin{equation}
H = -\frac{\nabla^2}{2\mr} + V_s(r) + \delta H \,.
\end{equation}
The static potential is given by
\begin{equation}
\label{eq:StatPot} 
V_s(r) = V_s^{(0)}(r) \left[ 1 + \sum\limits_{k=1}^{2} 
\left(\frac{\alpha_s}{4\pi}\right)^k a_k(r) \right]
\end{equation}
where, as mentioned above, $V_s^{(0)}(r) = -C_F \alpha_s/r$, 
is the leading order potential or Coulomb-like potential that goes 
directly in the Schr\"odinger equation. The ${\cal O}(\alpha_{s})$ 
and ${\cal O}(\alpha_{s}^{2})$ radiative corrections to the LO 
static potential are (the constants shown herein can be found, {\it 
e.g.}, in Appendix~C of Ref.~\cite{Pineda:2011dg}):
\begin{align}
\label{eq:radiative1}
a_1(\nu,r) &= a_1 + 2\beta_0 \ln{\nu \e{\gamma_E} r} \,, \\
\label{eq:radiative2}
a_2(\nu,r) &= a_2 + \frac{\pi^2}{3} \beta_0^2 + (4a_1 \beta_0 + 2\beta_1) 
\ln{\nu \e{\gamma_E} r} + 4\beta_0^2 \lnSquare{\nu \e{\gamma_E} r} \,.
\end{align}
The ${\cal O}(\alpha_{s})$ term was computed in Ref.~\cite{Fischler:1977yf} and 
the ${\cal O}(\alpha_s^2)$ in Ref.~\cite{Schroder:1998vy}. The static potential 
is known up to order ${\cal O}(\alpha_{s}^{4})$ with the ${\cal O}(\alpha_s^3)$ 
radiative correction computed in Refs.~\cite{Brambilla:1999qa, Kniehl:1999ud,
Anzai:2009tm, Smirnov:2009fh}. However, already ${\cal O}(\alpha_{s}^{3})$ terms 
would give a contribution to the E1 decay rate that goes beyond present precision.

The term $\delta H$ encodes the relativistic corrections which can be organized 
as an expansion in the inverse of the heavy quark mass, $m$. At the order we 
are interested in, such expansion covers all the $1/m$ and $1/m^2$ potentials 
and, at order $1/m^{3}$, the first relativistic correction to the kinetic 
energy:
\begin{equation}
\label{eq:deltaH} 
\delta H = -\frac{\nabla^4}{4 m^3} + \frac{V^{(1)}}{m} + 
\frac{V_{SI}^{(2)}}{m^2} + \frac{V_{SD}^{(2)}}{m^2} \,.
\end{equation}
At order $1/m^{2}$, we can split the contributions into spin-independent 
(SI) and spin-dependent (SD) terms~\cite{Brambilla:2004jw}
\begin{align}
V_{SI}^{(2)}(r) &= V_r^{(2)}(r) + \frac{1}{2} \lbrace 
V_{p^2}^{(2)}(r),-\nabla^2 \rbrace + V_{L^2}^{(2)}(r)\, \vec{L}^2 \,, \\
V_{SD}^{(2)}(r) &= V_{LS}^{(2)}(r)\, \vec{L} \cdot \vec{S} + V_{S^2}^{(2)}(r)\, 
\vec{S}^2 + V_{S_{12}}^{(2)}(r)\, S_{12} \,,
\end{align}
where $\vec{S}=\vec{S}_{1}+\vec{S}_{2}=(\vec{\sigma}_{1}+\vec{\sigma}_{2})/2$, 
$\vec{L}=\vec{r}\times\vec{p}$ and 
$S_{12}=3(\hat{r}\cdot\vec{\sigma}_{1})(\hat{r}\cdot\vec{\sigma}_{2})-\vec{
\sigma}_{1}\cdot\vec{\sigma}_{2} $ are, respectively, the total spin, 
total orbital angular momentum and tensor operators acting on the system. In 
the weak-coupling case, the above potentials read at leading (non-vanishing) 
order in perturbation theory
\begin{eqnarray}
V^{(1)}(r) = -\frac{C_F C_A \alpha_s^2}{2 r^2} \,, 
& V_r^{(2)}(r) = \pi C_F \alpha_s \delta^{(3)}(\vec{r}) \,, 
& V_{p^2}^{(2)}(r) = -\frac{C_F \alpha_s}{r} \,, \quad V_{L^2}^{(2)}(r) = 
\frac{C_F \alpha_s}{2 r^3} \,, \\
V_{LS}^{(2)}(r) = \frac{3 C_F \alpha_s}{2 r^3} \,,
& V_{S^2}^{(2)}(r) = \frac{4\pi C_F \alpha_s}{3} \delta^{(3)}(\vec{r}) \,, 
& V_{S_{12}}^{(2)}(r) = \frac{C_F \alpha_s}{4 r^3} \,.
\end{eqnarray}

We now make use of standard quantum mechanical perturbation theory and compute 
the first and second order correction, induced by a potential $V$, to a state 
$| n\l \rangle^{(0)} \equiv | n\l \rangle$. The second order correction to the 
wave function is only needed when the perturbation is given by the 
static potential proportional to the $a_{1}(\nu,r)$ term. The normalised 
corrected wave-function is
\begin{equation}
\label{eq:1st}
| n\l \rangle^{(1)} = \sum\limits_{n' \neq n \,,\, \l'} \frac{\langle n'\l' | V | 
n\l \rangle}{E_n^{(0)} - E_{n'}^{(0)}} | n'\l' \rangle \quad \left[ = 
\sum\limits_{n' \neq n \,,\, \l'} \frac{| n'\l' \rangle \langle n'\l' |}{E_n^{(0)} 
- E_{n'}^{(0)}} V | n\l \rangle \right] \,,
\end{equation}
for the first order, and
\begin{equation}
\label{eq:2nd}
| n\l \rangle^{(2)} = \!\!\!\! \sum\limits_{k_{1} \neq n \,,\, \l_1} \left[ 
\sum\limits_{k_{2} \neq n \,,\, \l_2} \!\!\!\! \frac{\langle k_{1}\l_1 | V | 
k_{2}\l_2 \rangle \langle k_{2}\l_2 | V | n\l \rangle}{(E_{n} - E_{k_{1}}) (E_{n} 
- E_{k_{2}})} \! - \! \frac{\langle k_{1}\l_1 | V | n\l \rangle \langle n\l | V | 
n\l \rangle}{(E_{n} - E_{k_{1}})^{2}} \right] | k_{1}\l_1 \rangle - \frac{1}{2} \!
\sum\limits_{k_{2} \neq n \,,\, \l_2} \!\!\!\! \frac{|\langle k_{2}\l_2 | V | 
n\l \rangle|^2}{(E_{n} - E_{k_{2}})^{2}} | n\l \rangle \,,
\end{equation}
for the second one.

As one can see in Eq.~(\ref{eq:1st}), a particular re-arrangement of the terms 
allows us to have a {\it key} expression that can be re-written as
\begin{equation}
\sum\limits_{n' \neq n \,,\, \l'} \! \frac{| n'\l' \rangle \langle n'\l' 
|}{E_n^{(0)} - E_{n'}^{(0)}} = \! \sum\limits_{n' \,,\, \l'} \! \frac{| n'\l' 
\rangle \langle n'\l' |}{E_n^{(0)} - E_{n'}^{(0)}} - \!\!\!\! \sum\limits_{n'=n 
\,,\, \l'} \! \frac{| n'\l' \rangle \langle n'\l' |}{E_n^{(0)} - E_{n'}^{(0)}} = 
\!\!\!\!\! \lim\limits_{ ~~ E \to E_n^{(0)}} \left( \frac{\1}{E - H} - 
\frac{\mathcal{P}(n)}{E - E_n^{(0)}} \right) \equiv \frac{1}{(E_n - H)'} \,.
\end{equation}
This will allow us to compute expectation values of an arbitrary operator $\O$, 
via (note that, for the sake of simplicity, only final state corrections are 
shown here but the same corrections affect also the initial state):
\begin{equation}
\begin{split}
\label{eq:1st2nd}
\langle n'\l' | \O | n\l \rangle^{(1)} &= \langle n'\l' | \O \frac{1}{(E_n - 
H)'} V | n\l \rangle \\
&
= \int \d^3 r_1 \, \d^3 r_2 \, \psi_{n'\l'}^*(\vec{r}_2) \, \O(\vec{r}_2) \, 
G'_n(\vec{r}_2,\vec{r}_1) \, V(\vec{r}_1) \, \psi_{n\l}(\vec{r}_1) \,,
\end{split}
\end{equation}
for the first order, and
\begin{equation}
\begin{split}
\label{eq:2nd2nd}
\langle n'\l' | \O | n\l \rangle^{(2)} &= \langle n'\l' | \O \frac{1}{(E_n - 
H)'} V \frac{1}{(E_n - H)'} V | n\l \rangle \\
& 
- \langle n\l | V | n\l \rangle \langle n'\l' | \O 
\frac{1}{(E_n - H)'} \1 \frac{1}{(E_n - H)'} V | n\l \rangle \\
&
- \frac{1}{2} \langle n'\l' | \O | n\l \rangle \langle n\l | V \frac{1}{(E_n - 
H)'} \1 \frac{1}{(E_n - H)'} V | n\l \rangle \\
&
= \int \d^3 r_1 \, \d^3 r_2 \, \d^3 r_3 \, 
\psi_{n'\l'}^*(\vec{r}_3) \, \O(\vec{r}_3) \, G'_n(\vec{r}_3,\vec{r}_2) \, 
V(\vec{r}_2) \, G'_n(\vec{r}_2,\vec{r}_1) \, V(\vec{r}_1) \psi_{n\l}(\vec{r}_1) \\
& 
- \delta E_V^{(1)} \times \int \d^3 r_1 \, \d^3 r_2 \, \d^3 
r_3 \, \psi_{n'\l'}^*(\vec{r}_3) \, \O(\vec{r}_3) \, G'_n(\vec{r}_3,\vec{r}_2) \, 
G'_n(\vec{r}_2,\vec{r}_1) \, V(\vec{r}_1) \psi_{n\l}(\vec{r}_1) \\
&
- \frac{1}{2} \int \d^3 r \, \psi_{n'\l'}^*(\vec{r}) \, \O(\vec{r}) \, 
\psi_{n\l}(\vec{r}) \, \times \\
& 
\times \int \d^3 r_1 \, \d^3 r_2 \, \d^3 r_3 \, 
\psi_{n'\l'}^*(\vec{r}_3) \, V(\vec{r}_3) \, G'_n(\vec{r}_3,\vec{r}_2) \, 
G'_n(\vec{r}_2,\vec{r}_1) \, V(\vec{r}_1) \psi_{n\l}(\vec{r}_1) \,,
\end{split}
\end{equation}
for the second order. The term $\delta E_V^{(1)}$ in Eq.~(\ref{eq:2nd2nd}) is 
the first order correction to the energy induced by a potential $V$: $\delta 
E_V^{(1)} \equiv \int \d^3 r \, \psi_{n'\l'm'}^*(\vec{r}\,) \, V(\vec{r}\,) \, 
\psi_{n\l m}(\vec{r}\,)$; and $G^{\prime}_n(\vec{r}_1,\vec{r}_2)$ has the 
following expression
\begin{equation}
G'_n(\vec{r}_1,\vec{r}_2) \equiv (-1) \times \!\!\! \lim_{ ~ E \to E_n} 
\left(G(\vec{r}_1,\vec{r}_2,E) - \sum\limits_{\l=0}^\infty \frac{|\psi_{n\l}|^2}{E 
- E_n}\right) \,,
\end{equation}
where $G(\vec{r}_1,\vec{r}_2,E)$ is the Coulomb Green function
\begin{align}
G(\vec{r}_1,\vec{r}_2,E) &= \sum\limits_{\l=0}^\infty \frac{2\l+1}{4\pi} 
P_\l(\hat{r}_1 \cdot \hat{r}_2) G_\l(r_1,r_2) \,, \\
G_\l(r_1,r_2) &= \sum\limits_{\nu=\l+1}^\infty \mr a^2 
\left(\frac{\nu^4}{\lambda}\right) 
\frac{R_{\nu\l}(\r{\lambda}{1})R_{\nu\l}(\r{\lambda}{2})}{\nu-\lambda} \,.
\end{align}
in which we have defined $E \equiv -\frac{m_{r} C_{F}^{2} 
\alpha_s^{2}}{2\lambda^{2}}$.\footnote{In order to perform the computation it is 
specially useful to use such expression, because for $\lambda = \frac{n}{\sqrt{1 - 
\epsilon}}$, we have $E = E_{n} (1 - \epsilon)$ and $E \to E_n$ for $\epsilon \to 
0$.}


\subsubsection{Corrections due to higher order Fock states}

\begin{figure}[!t]
\centerline{%
\includegraphics[clip,width=0.44\textwidth]
{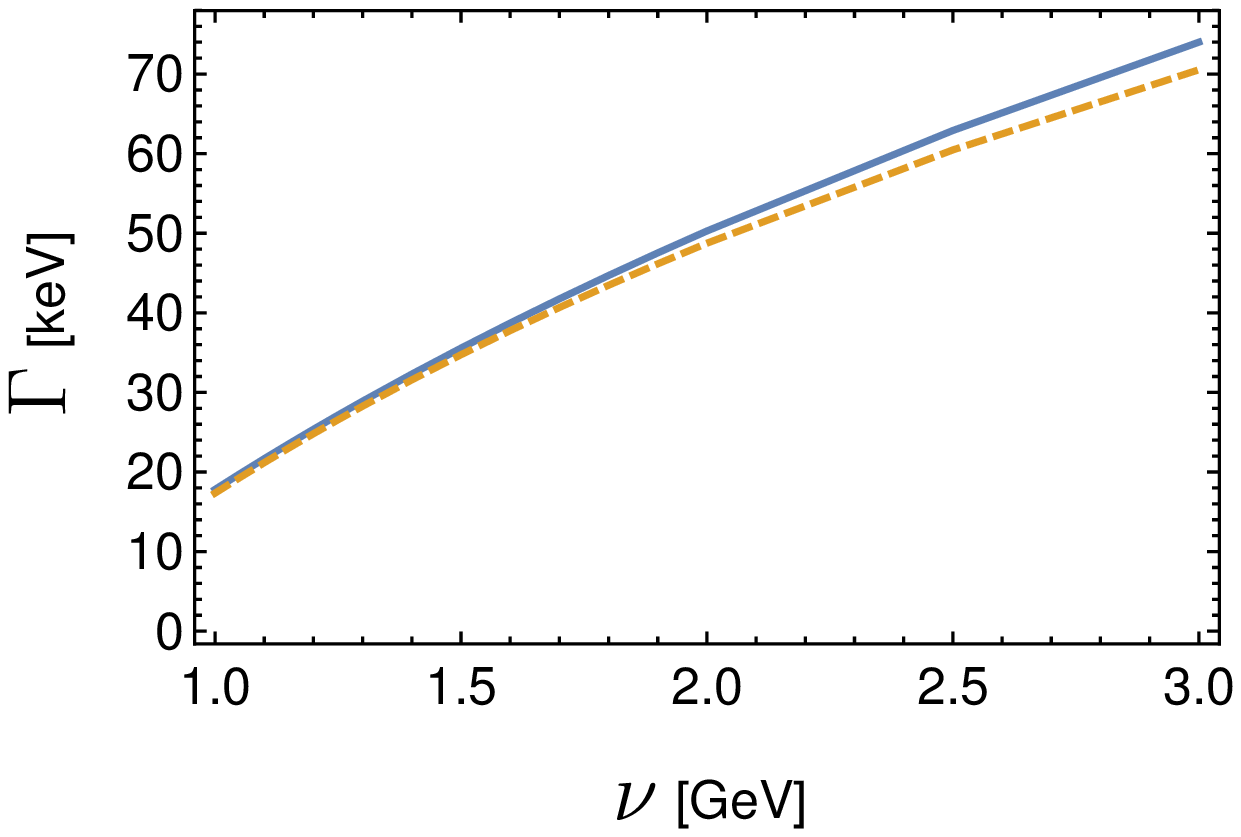}
\hspace*{0.50cm}
\includegraphics[clip,width=0.44\textwidth]
{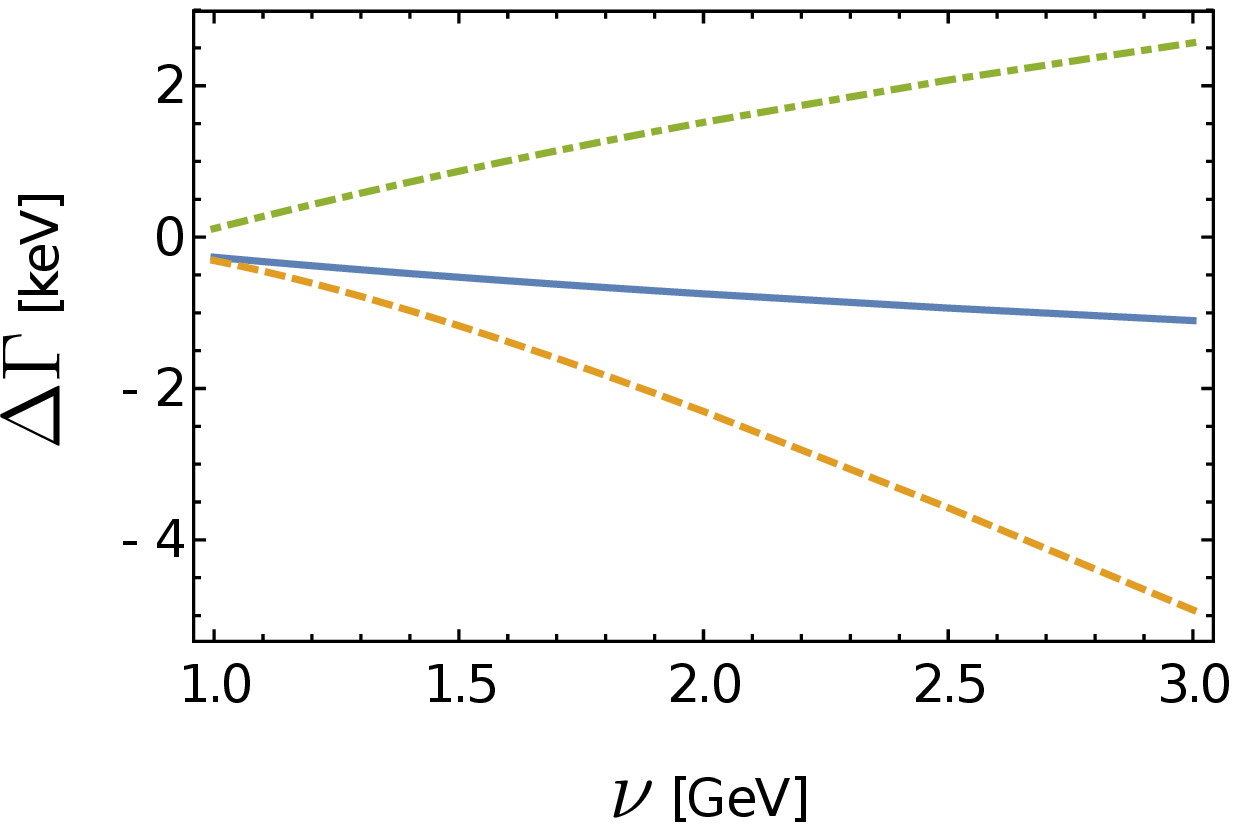}
}
\caption{\label{fig1} 
{\it Left panel} -- Comparison between the LO decay width (solid blue curve) 
and its relativistic correction (dashed orange curve) due to higher order 
electromagnetic transition operators that appear in the pNRQCD Lagrangian as a 
function of the renormalization scale $\nu$.
{\it Right panel} -- Relativistic contributions appearing in 
Eq.~\eqref{eq:FullDecayWidth}: The first $\Gamma^{(0)}_{E1} \times 
(-k_\gamma/(6m))$ (solid blue), the second $\Gamma^{(0)}_{E1} \times 
(-k_\gamma^2/(60) \times I_5^{(0)}/I_3^{(0)})$ (dashed orange) and the third 
$\Gamma^{(0)}_{E1} \times (\text{second line of Eq.~\eqref{eq:FullDecayWidth}})$ 
(dot-dashed green).}
\end{figure}

The weakly coupled quarkonia may also get corrections from the coupling of the 
heavy quark-antiquark pair to other low-energy degrees of freedom. In 
particular, the leading order electromagnetic dipole transition may get a
correction from diagrams (see Fig. 8 in \cite{Brambilla:2012be}) in which a 
singlet state is coupled to an octet state due to the emission and re-absorption 
of an ultrasoft gluon. These diagrams come from terms of the pNRQCD Lagrangian 
like~\cite{Brambilla:2004jw}
\begin{equation}
\Delta{\cal L} = V_{A} \left(O^{\dagger} \vec{r}\cdot g\vec{E} S + S^{\dagger} 
\vec{r}\cdot g\vec{E} O \right) \,,
\end{equation}
where $S=S 1_{c}/\sqrt{N_{c}}$ is a quark-antiquark field that transforms as a 
singlet under $SU(3)_{\rm c}$ and $U(1)_{\rm em}$, $O = \sqrt{2} O^{a}T^{a}$ is 
a quark-antiquark field which transforms as an octet under $SU(3)_{\rm c}$ and 
as a singlet under $U(1)_{\rm em}$, and $\vec{E}$ is the chromo-electric field.

The first two diagrams in Fig. 8 of \cite{Brambilla:2012be} correspond to the 
renormalization of the initial and final wave function. The diagrams 2, 3a and 
3b account for the correction of the initial and final wave functions due to 
the presence of octet states. The diagram 4 represents an electric dipole 
transition mediated by the intermediate octet state. According to the power 
counting, the first two diagrams contribute to relative order $\Lambda_{\rm 
QCD}^{2}/(mv)^{2}$ whereas the remaining ones scales as $\Lambda_{\rm 
QCD}^{3}/(mv^{2})/(mv)^2$. We shall not consider these contributions herein 
because in the strict weak-coupling regime, $E\sim mv^{2} \gg \Lambda_{\rm QCD}$, 
one can argue that they should be negligible. 

It is noteworthy that, in contrast to the E1 transitions, the colour-octet 
contributions for allowed M1 transitions cancel~\cite{Brambilla:2005zw}. This is 
a consequence of the fact that the magnetic dipole operator behaves as an identity 
operator in position space.


\section{Results}
\label{sec:Results}

\begin{figure}[!t]
\centerline{%
\includegraphics[clip,width=0.32\textwidth]{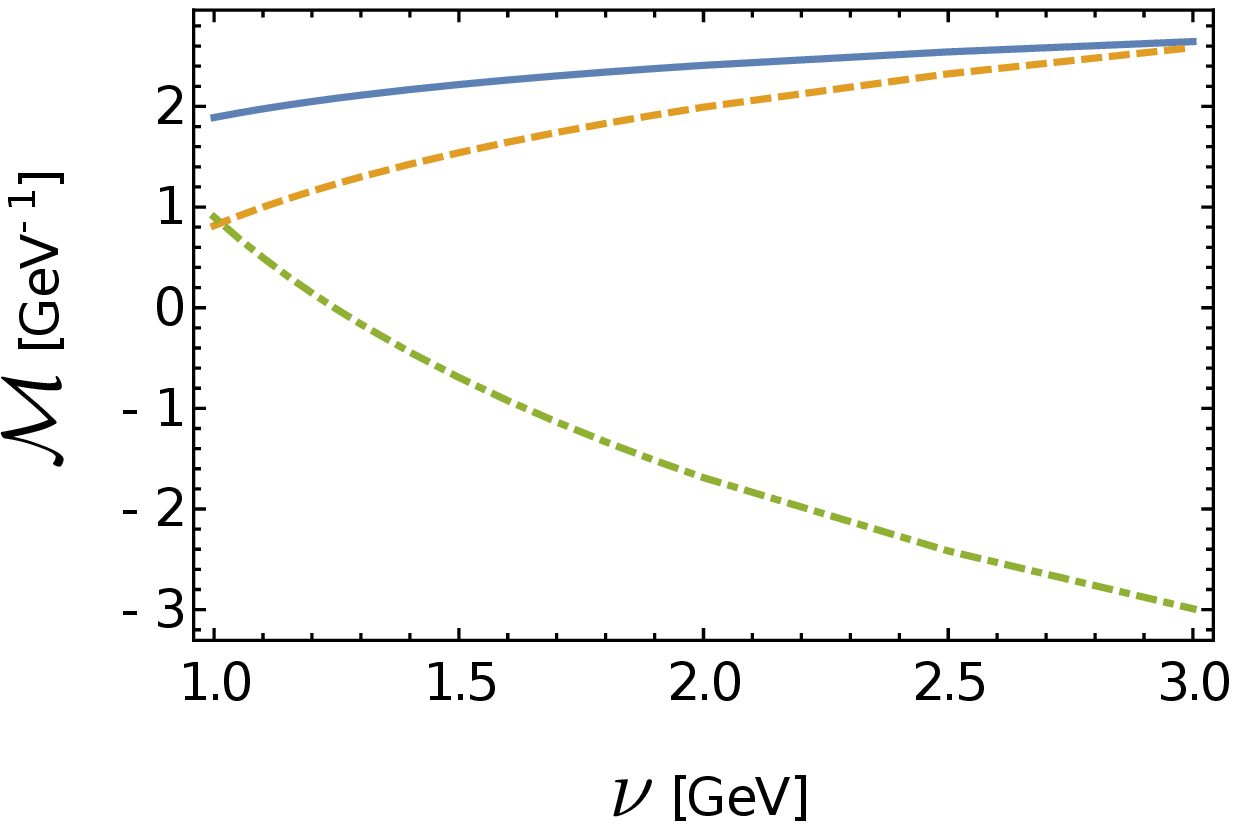}
\includegraphics[clip,width=0.32\textwidth]{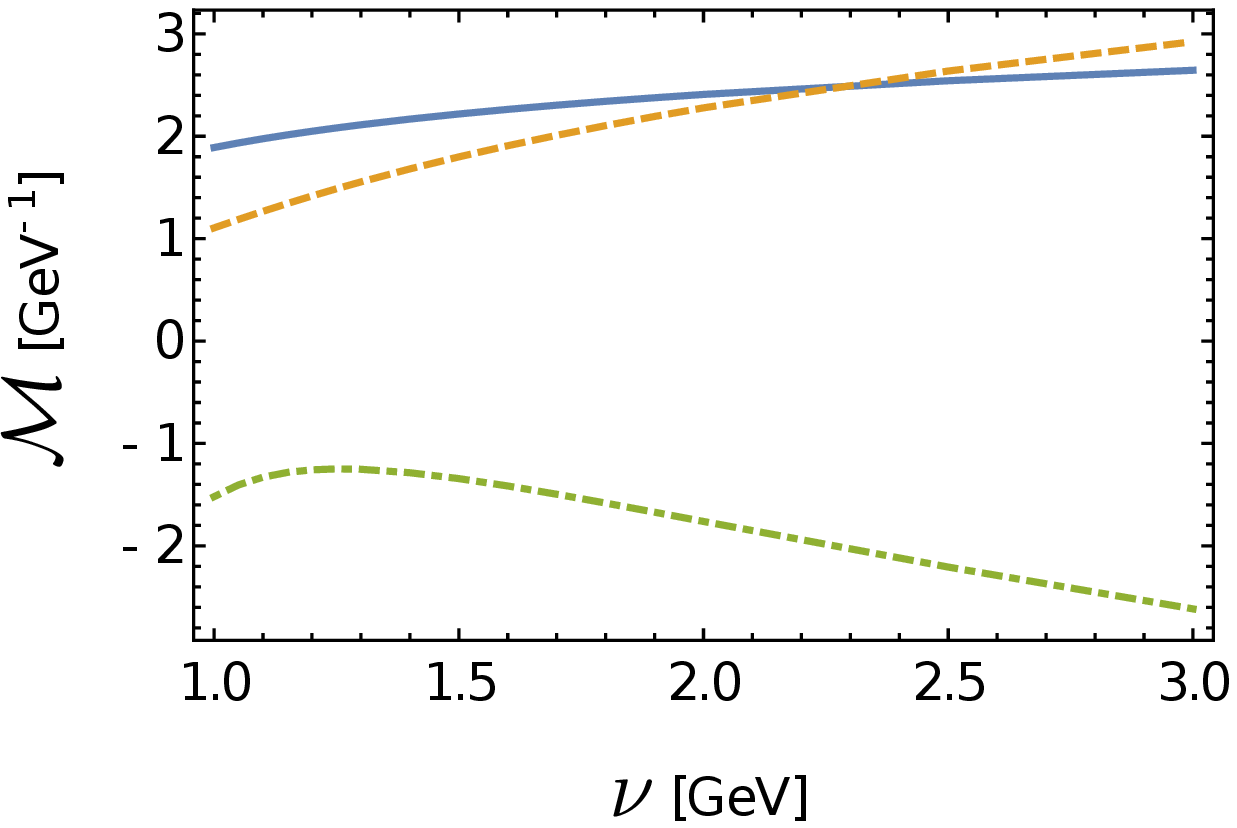}
\includegraphics[clip,width=0.32\textwidth]{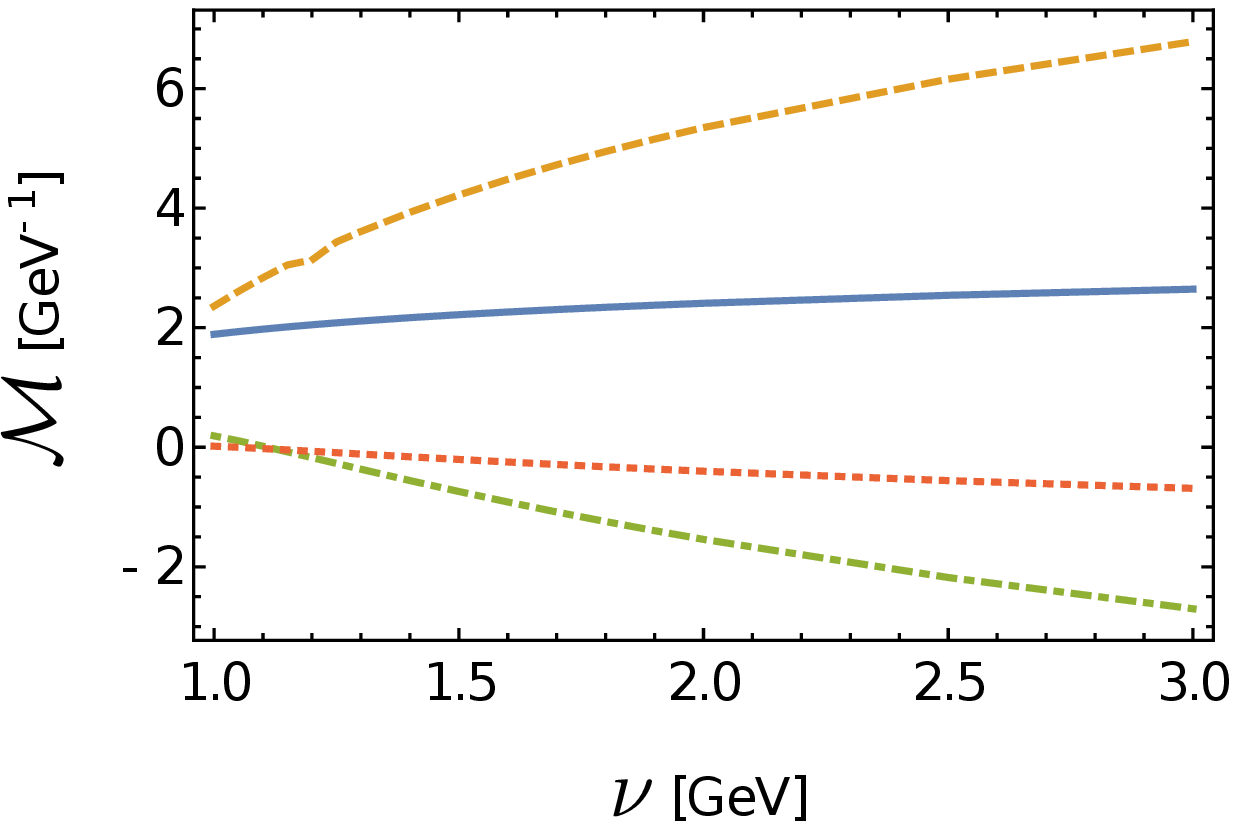}}
\vspace*{0.20cm}
\centerline{%
\includegraphics[clip,width=0.32\textwidth]{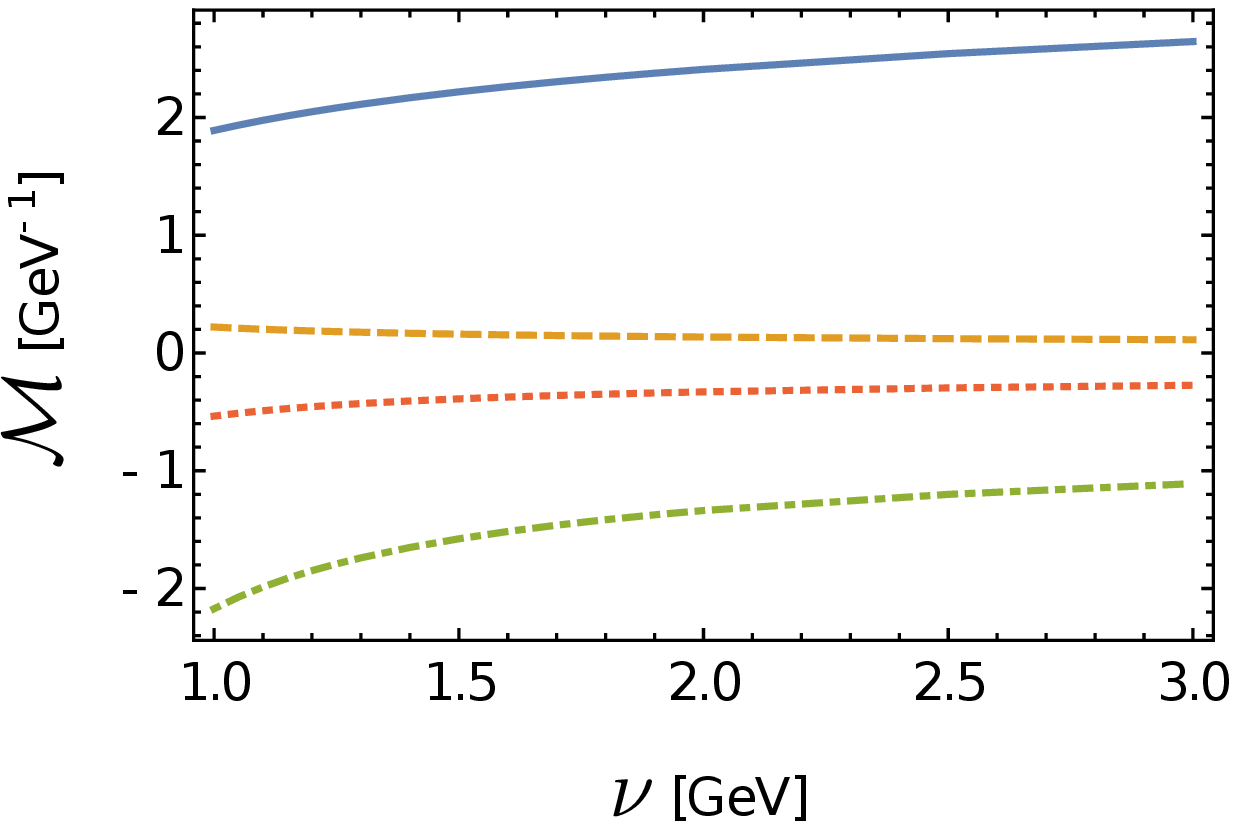}
\includegraphics[clip,width=0.32\textwidth]{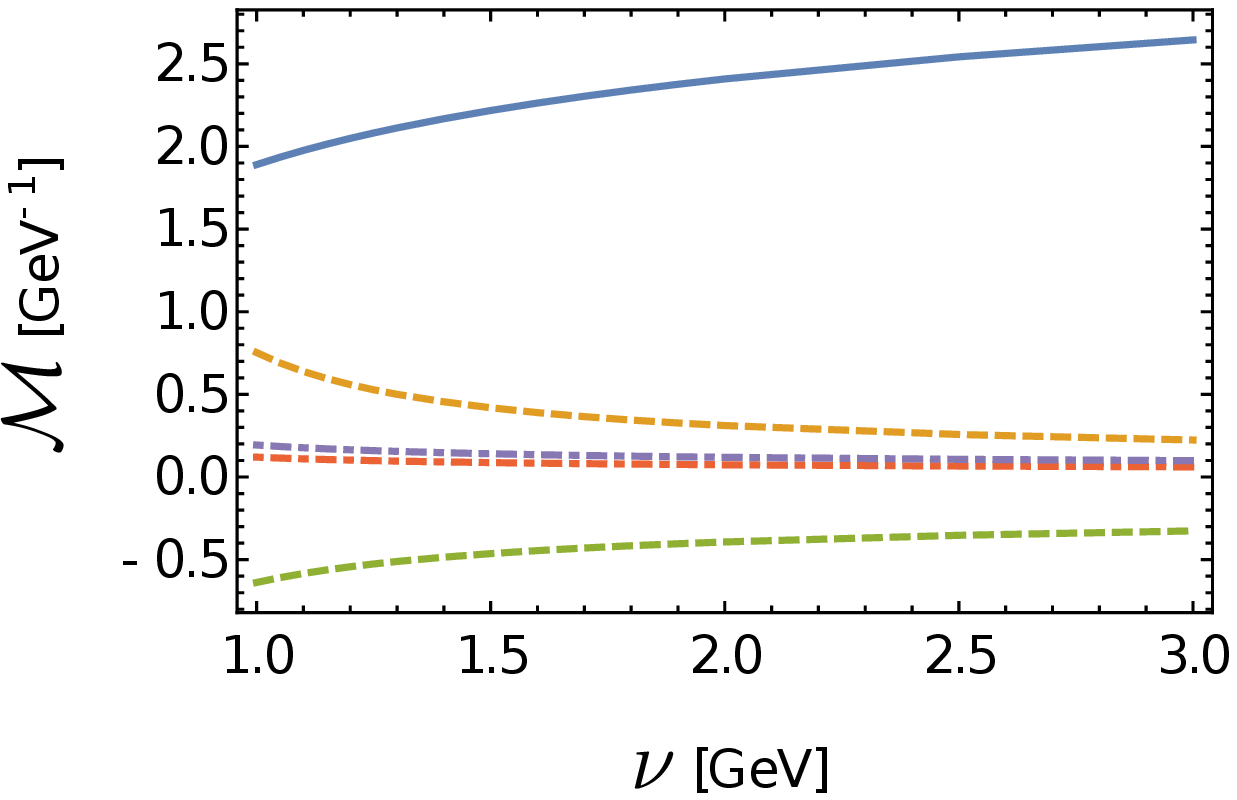}
\includegraphics[clip,width=0.32\textwidth]{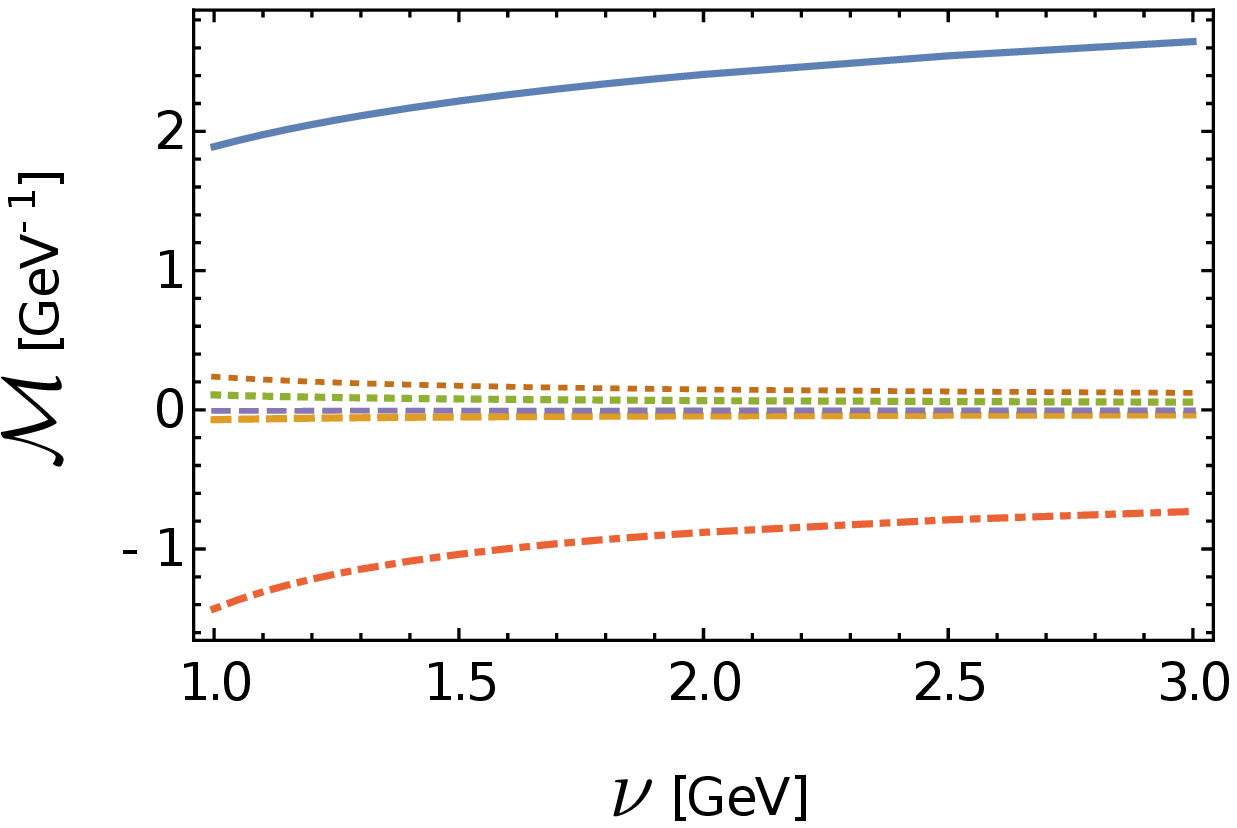}}
\caption{\label{fig2} Comparison of the LO transition matrix element (solid 
blue curve) with respect the ones coming from corrections due to higher order 
potentials.
{\it Upper-left panel} -- First order correction to the initial (dashed yellow) 
and final (dot-dashed green) wave functions due to the $a_{1}(\nu,r)$ term.
{\it Upper-middle panel} -- First order correction to the initial (dashed 
yellow) and final (dot-dashed green) wave functions due to the $a_{2}(\nu,r)$.
{\it Upper-right panel} -- Second order correction to the initial (dashed 
yellow) and final (dot-dashed green) wave functions, and first order correction 
to both initial and final (dotted red) due to the $a_{1}(\nu,r)$ term.
{\it Lower-left panel} -- First order correction to the initial (dashed yellow) 
and final (dot-dashed green) wave functions due to the $V^{(1)}$, and first 
order correction to the final (dotted red) wave functions due to the 
$V_{r}^{(2)}$.
{\it Lower-middle panel} -- First order correction to the initial (dashed 
yellow) and final (dashed green) wave functions due to the $p^{2}$ term; 
moreover, first order correction to the initial (dot-dashed red) and final 
(dot-dashed purple) wave functions due to the kinetic $p^{4}$ term.
{\it Lower-right panel} -- Remaining contributions where the most important one 
(dot-dashed red) is the first order correction to the final wave function due 
to the $V_{S^2}^{(2)}$.
}
\end{figure}

We discuss in detail our theoretical result for the $\chi_{b1}(1P)\to 
\Upsilon(1S)\gamma$ reaction, but a similar analysis has been performed for the 
transitions $\chi_{bJ}(1P)\to \Upsilon(1S)\gamma$ with $J=0,\,2$. The mean value 
for the decay width and an estimate of its theoretical error will be given at the 
end of this Section for all transitions.

Figure~\ref{fig1} shows the $\chi_{b1}(1P)\to \Upsilon(1S)\gamma$ LO decay rate 
and its relativistic correction due to higher order electromagnetic 
interactions that appear in the pNRQCD Lagrangian. In other words, we are 
analysing Eq.~\eqref{eq:FullDecayWidth} without the factor $R^{S=1}(J=1)$. As 
one can see in the left panel of Fig.~\ref{fig1}, these ${\cal O}(v^{2})$ 
corrections to the LO decay rate are very small, $\sim\!5\%$ at most. The right 
panel of the same figure displays the different contributions (with their 
relative sign) showing that the dominant one is the term proportional to 
$I_{5}^{(0)}(21\to10)$ in the expression of the decay rate, 
Eq.~\eqref{eq:FullDecayWidth}. An interesting feature shown in Fig.~\ref{fig1} 
is the substantial dependence of the result on the renormalization scale $\nu$. 
The decay width changes from $17\,{\rm keV}$ to $74\,{\rm keV}$ when the 
renormalization scale $\nu$ is varied within the range of $1$ to $3\,{\rm GeV}$. 
This range encompasses the typical momentum transfer in the  bottomonium system, 
still being consistent with perturbation theory.

Let us focus now our attention to the computation of the wave function 
corrections due to higher order potentials, which are encoded in the factor 
$R^{S=1}(J=1)$ of Eq.~\eqref{eq:FullDecayWidth}. The upper panels of 
Fig.~\ref{fig2} show the matrix elements correcting the E1 decay rate up to 
${\cal O}(v^{2})$ and coming from the radiative corrections to the static 
potential, Eq.~\eqref{eq:StatPot}. The left and middle panels refer to the first 
order initial and final wave function corrections coming from $a_{1}(\nu,r)$ 
and $a_{2}(\nu,r)$, respectively. The right panel refers to the second order 
correction due to the $a_{1}(\nu,r)$ term of the static potential. Amongst the 
features shown by the panels, the following are of particular interest: (i) the 
matrix elements clearly exceed the value of the LO one. (ii) The matrix elements 
depend quite dramatically on the scale $\nu$, especially for small $\nu$; in 
some sense, we expected such behaviour from the numerical analysis of the M1 
transitions in Refs.~\cite{Brambilla:2005zw, Pineda:2013lta}. (iii) The zero 
crossing in some of the matrix elements comes from the logarithms in 
\eqref{eq:radiative1} and \eqref{eq:radiative2}.

The lower panels of Fig.~\ref{fig2} show the remaining matrix element 
contributions coming from $\delta H$, Eq.~\eqref{eq:deltaH}. One can see that 
only few of them are relevant corrections to the LO decay rate. Moreover, the 
$\nu$-dependence of every matrix element is smaller than in the case of the 
radiative corrections.\footnote{The only two parameters in our approach are 
$m_{b}$ and $\alpha_s$. The value of the b-quark mass is fixed through the 
$\Upsilon(1S)$-mass and the running of $\alpha_s(\nu)$ is taken at 4-loop accuracy 
with three massless flavours using the Mathematica package 
RunDec~\cite{Chetyrkin:2000yt} and the starting value $\alpha_s(M_Z) = 0.118$.}

\begin{figure}[!t]
\centerline{%
\includegraphics[clip,width=0.44\textwidth]{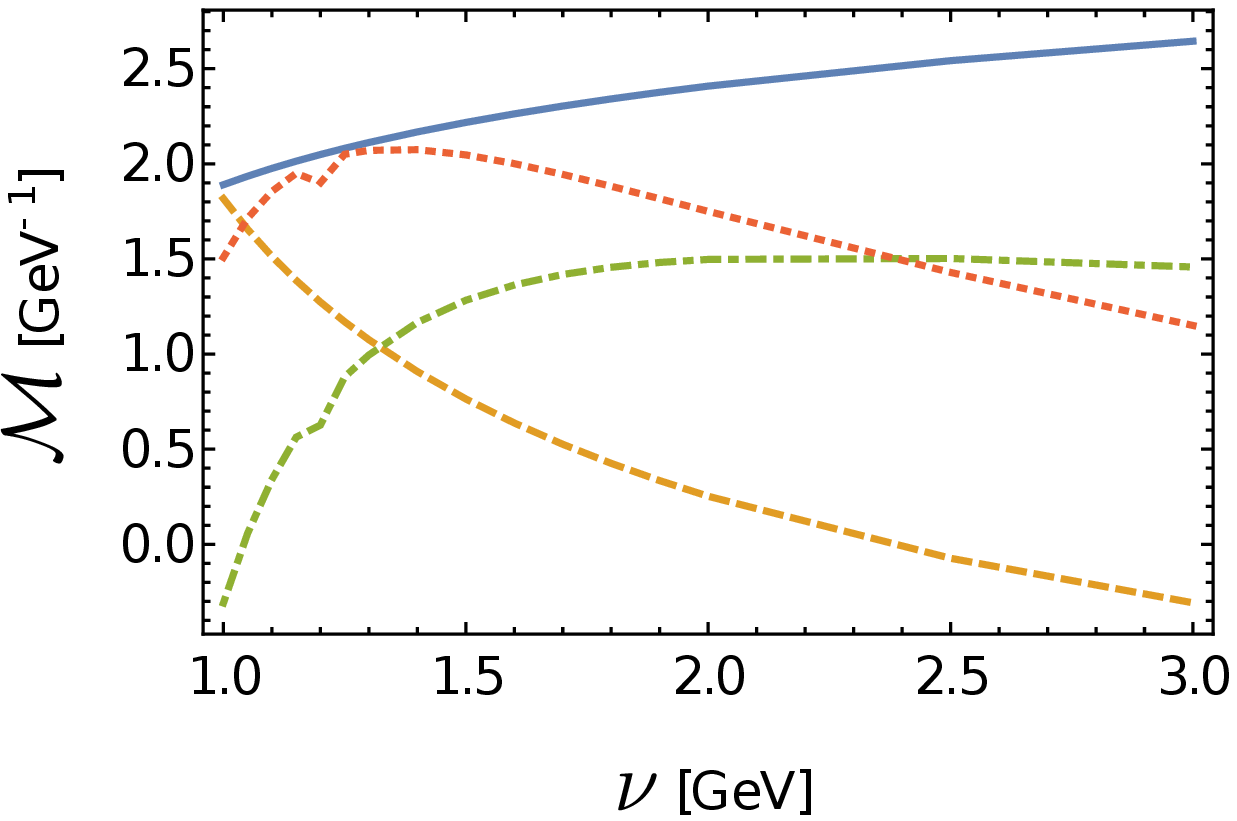}
\hspace*{0.50cm}
\includegraphics[clip,width=0.44\textwidth]{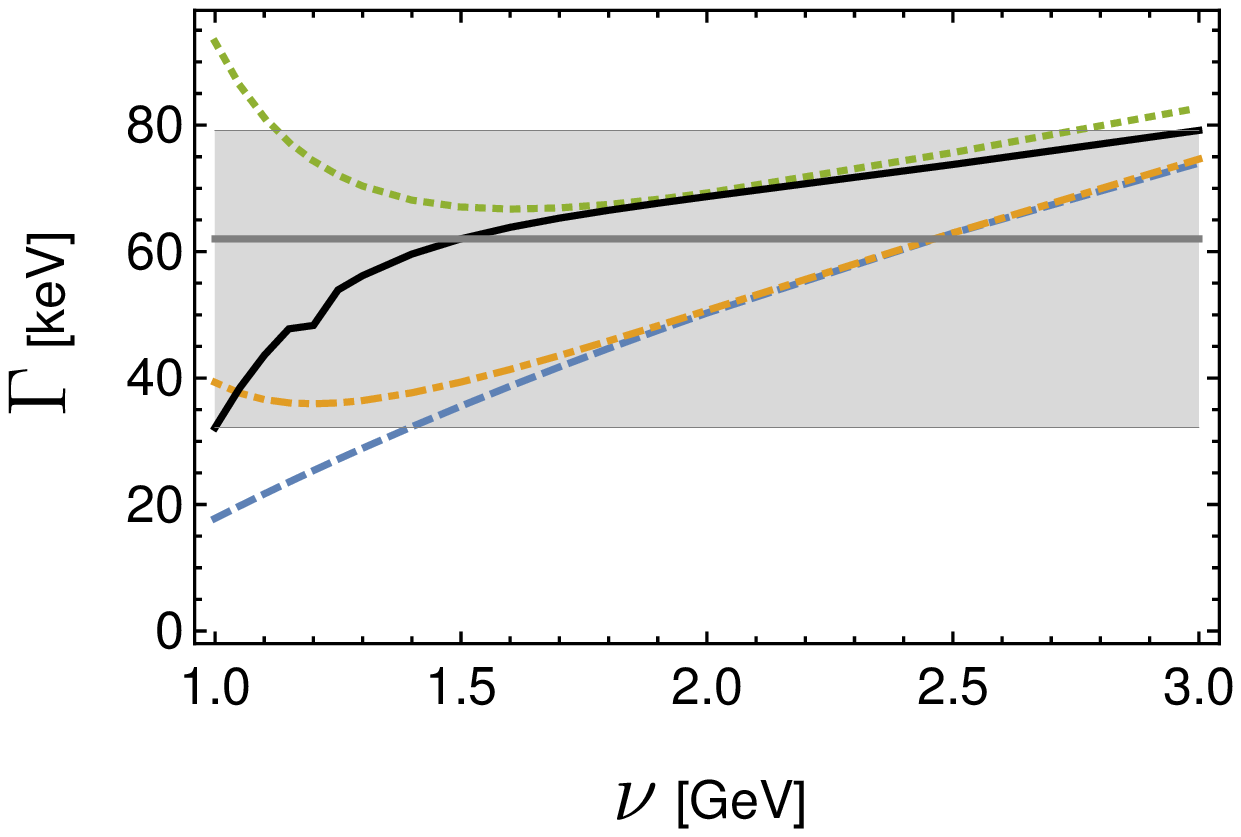}
}
\caption{\label{fig3} 
{\it Left panel} -- Matrix elements (with their relative signs) contributing 
to the reaction $\chi_{b1}(1P)\to \Upsilon(1S)\gamma$ at LO order (solid blue), 
NLO (dashed yellow), NNLO (dot-dashed green) and NLO+NNLO (dotted red).
{\it Right panel} -- Total decay width for the $\chi_{b1}(1P)\to 
\Upsilon(1S)\gamma$ reaction, the panel shows the LO (dashed blue), LO+NLO 
(dot-dashed yellow) and LO+NLO+NNLO (solid black) result. The green dotted curve 
is the LO+NLO+NNLO result but omitting the contributions coming from the radiative 
corrections to the static potential. The horizontal gray line is our final value 
for the decay width, taken at $\nu = 1.5$~GeV, and the gray band corresponds to 
the uncertainty ($44.23\% \, \hat{=} \pm 26.05$~keV for this transition).
}
\end{figure}

Summing all the contributions discussed in the paragraph above, the left panel 
of Fig.~\ref{fig3} shows the next-to-leading order (NLO), NNLO and NLO+NNLO 
matrix elements and compares them with the LO term. The most important features 
have been already mentioned: the subleading matrix elements are of the same 
order of magnitude than the leading one and the dependence with $\nu$ 
in the logs dominates the picture. In the right panel of Fig.~\ref{fig3}, we 
draw the decay rate associated with the $\chi_{b1}(1P)\to \Upsilon(1S)\gamma$ 
reaction at LO, NLO and NNLO. It is worth to remark that the NLO contribution is 
negligible at large-$\nu$ but multiplies by a factor of 2 the LO decay width at 
$\nu=1\,{\rm GeV}$. A big correction to the decay rate is due to the NNLO 
contribution. One can see in the Figure that the theoretical result depends 
slightly on the scale for $\nu\gtrsim1.75\,{\rm GeV}$, whereas the 
$\nu$-dependence is dramatic for lower values due to the logarithmic functions. 
This fact is demonstrated by the additional curve (dotted green) where we omitted 
the contributions coming from the radiative corrections to the static potential, 
hence set the $a_{1}(\nu,r)$ and $a_{2}(\nu,r)$ terms to zero. Note that the 
relativistic corrections to the leading order E1 transition operator are included 
in the NNLO curve.

Finally, our theoretical results for the decay rates of the transitions under 
consideration are obtained by choosing the value at $\nu = 1.5$~GeV, yielding:
\begin{align}
\Gamma(\chi_{b0}(1P)\to\Upsilon(1S)\gamma) &= \left(52^{+14}_{-24}
({\cal O}(v^{3}))\right)\,{\rm keV} \,, \\
\Gamma(\chi_{b1}(1P)\to\Upsilon(1S)\gamma) &= \left(62^{+17}_{-30} 
({\cal O}(v^{3}))\right)\,{\rm keV} \,, \\
\Gamma(\chi_{b2}(1P)\to\Upsilon(1S)\gamma) &= \left(64^{+18}_{-33}
({\cal O}(v^{3}))\right)\,{\rm keV} \,,
\end{align}
where we have chosen a very conservative error estimation that includes the total 
range of our final result, obtained by varying $\nu$ from (1-3)~GeV.


\section{Epilogue}
\label{sec:epilogue}

We have presented the first numerical determination of the decay rate 
$\chi_{bJ}(1P)\to \Upsilon(1S)\gamma$ with $J=0,\,1,\,2$ within potential NRQCD. 
We have assumed that the momentum scale of the heavy quarkonium involved lies in 
the strict weak-coupling regime of pNRQCD and non-perturbative effects are 
suppressed, such that a full perturbative calculation can be performed. 
Relativistic corrections of relative order $v^{2}$ to the LO decay rate are 
included. The analysis separates those contributions that account for the higher 
order electromagnetic interaction terms in the pNRQCD Lagrangian and those that 
account for quarkonium state corrections due to higher order potentials and 
transitions between singlets and octets.


\begin{acknowledgement}
S.S. and J.S. thank N. Brambilla and A. Vairo for collaboration and supervision 
on the work presented here and C. Peset, A. Pineda, Y. Sumino and Y. Kiyo for 
numerous informative discussions.
S.S. expresses his gratitude to the Physik-Department of the Technische 
Universit\"at M\"unchen whose support helped his participation in the 
Conference. J.S. acknowledges the financial support from the Alexander von 
Humboldt Foundation.
\end{acknowledgement}


\bibliography{STEINBEISSER_Sebastian_CONF12}

\end{document}